# Bayesian $\mathcal{F}$-statistic-based parameter estimation of continuous gravitational waves from known pulsars

A. Ashok[1,2] P. B. Covas[1,2] R. Prix[1,2] and M. A. Papa[1,2]

[1]*Max Planck Institute for Gravitational Physics (Albert Einstein Institute), D-30167 Hannover, Germany*
[2]*Leibniz Universität Hannover, D-30167 Hannover, Germany*

We present a new method and implementation to obtain Bayesian posteriors on the amplitude parameters $\{h_0, \cos\iota, \psi, \phi_0\}$ of continuous gravitational waves emitted by known pulsars. This approach leverages the well-established $\mathcal{F}$-statistic framework and software. We further explore the benefits of employing a likelihood function that is analytically marginalized over $\phi_0$, which avoids signal degeneracy problems in the $\psi$-$\phi_0$ subspace. The method is tested on simulated signals, hardware injections in Advanced-LIGO detector data, and by performing percentile-percentile self-consistency tests of the posteriors via Monte-Carlo simulations. We apply our methodology to PSR J1526-2744, a recently discovered millisecond pulsar. We find no evidence for a signal and obtain a Bayesian upper limit $h_0^{95\%}$ on the gravitational-wave amplitude of approximately $7 \times 10^{-27}$, comparable with a previous frequentist upper limit.

## I. INTRODUCTION

Continuous gravitational waves (CWs) are long-lasting periodic gravitational wave signals the detection of which is one of the goals of gravitational wave astronomy.

The simplest way to produce continuous gravitational waves that could be detected by the current generation of detectors is through a varying mass quadrupole moment in a fast-rotating neutron star. In the absence of precession, the signal is expected at twice the rotation frequency and with twice the rotational spin-down of the star. The position and orientation of the neutron star influence how the gravitational wave couples to the detector, and the position also determines the observed gravitational-wave phase, through the Doppler effect. All in all the signal is described by $p + 4$ parameters: the $p$ so-called phase-evolution parameters (frequency and its first $k$ derivatives, sky position and binary orbital parameters if applicable) and the amplitude parameters $\{h_0, \cos\iota, \psi, \phi_0\}$—intrinsic amplitude, orientation, polarization angle, and initial phase, respectively.

There are three broad classes of CW searches, depending on the amount of knowledge available on the source. All-sky searches [1–6] assume no information about the sources and search over a broad signal parameter space. Directed searches [7–13] focus on objects with known sky position but have limited or no knowledge on their spin parameters. Targeted searches [14–17] use electromagnetic observations of pulsars to accurately infer the gravitational-wave phase-evolution parameters.

Owing to the massive reduction in parameter-space size compared to wide-parameter-space searches, a targeted search using a fully coherent combination of all the data, leading to the maximum possible sensitivity, is possible [14,16,18,19]. Narrow-band searches around expected signal parameters can typically also still be performed at nearly maximum sensitivity [16,18].

Since the presence of a neutron star is assured and its rotational frequency and spin-down are known, a null measurement is directly informative about the gravitational-wave emission of the source. Targeted searches of known sources are, therefore, a crucial class of CW searches.

Previously, only a single Bayesian method and implementation existed for amplitude-parameter estimation on known pulsars [20], often referred to as the *Time Domain method* or *Heterodyne method*. This method has been successfully used for targeted searches for a long time [14,16,21–26].

In this paper, we introduce a new expression and implementation of the CW signal likelihood function, based on the well-established $\mathcal{F}$-statistic framework. Combining this likelihood with standard stochastic (Markov Chain Monte Carlo (MCMC) and nested) sampling methods allows us to perform the Bayesian parameter estimation. In principle, this approach can be used for any subspace of the full CW parameter space, but in this first study we focus on *targeted* searches, where all phase-evolution parameters of the source are assumed to be known, and the posterior is computed over the unknown amplitude parameters only (i.e., amplitude $h_0$ and orientation angles $\{\iota, \psi, \phi_0\}$ of the source). As an incidental benefit, this avoids convergence difficulties for the samplers that can arise if the parameter space is too large; e.g., see [27].

The paper is organized as follows. In Sec. II, we describe the continuous gravitational wave signal model. In Sec. III, we derive the $\mathcal{F}$-statistic-based likelihood function, describe its software implementation and discuss two tests to validate the method. Section IV introduces and tests a likelihood function that is analytically marginalized over the initial-phase parameter $\phi_0$. Section V illustrates the application of the method to the hardware injections in Advanced LIGO data. In Sec. VI, we apply this method to perform parameter estimation on a putative CW signal from PSR J1526-2744 and obtain a Bayesian upper limit on $h_0$ from this posterior. Section VII summarizes the method and the results and discusses possible future work.

## II. SIGNAL MODEL

We assume that the signal is a nearly monochromatic CW of the form described in Sec. II of [28]. The signal strain in the detector has the form

$$s(t) = F_+(t;\alpha,\delta,\psi)h_+(t) + F_\times(t;\alpha,\delta,\psi)h_\times(t), \quad (1)$$

where "+" and "×" indicate the two gravitational-wave polarizations, and $F_+(t;\alpha,\delta,\psi)$ and $F_\times(t;\alpha,\delta,\psi)$ are the detector antenna-pattern functions. These depend on the relative orientation between the detector and the source as a function of time $t$, and on the sky position $(\alpha,\delta)$ of the source and the polarization angle $\psi$. The two waveforms $h_+(t)$ and $h_\times(t)$ are given by

$$h_+(t) = A_+ \cos\phi(t),$$
$$h_\times(t) = A_\times \sin\phi(t), \quad (2)$$

with the two polarizations amplitudes $A_{+,\times}$ expressible as

$$A_+ = \frac{1}{2}h_0(1+\cos^2\iota), \qquad A_\times = h_0 \cos\iota, \quad (3)$$

in terms of the overall amplitude $h_0$ and the inclination angle $\iota$ between the neutron star angular momentum and the line of sight. The signal phase $\phi(t)$ in Eq. (2) in the detector frame at time $t$ depends on the signal frequency $f$ and its derivatives $f^{(k)}$ (at some reference time), as well as the source sky position, and—if the neutron star is in a binary system—the binary orbital parameters $b$. As already anticipated in the previous section, these are collectively referred to as the *phase-evolution* parameters $\lambda \equiv \{\alpha,\delta,f,\dot{f},\ldots,b\}$.

As shown in [28] the signal amplitude parameters $\{h_0, \cos\iota, \psi, \phi_0\}$ can be reparametrized into a set of four amplitude coordinates $\mathcal{A}^\mu$, defined as

$$\mathcal{A}^1 \equiv A_+ \cos\phi_0 \cos 2\psi - A_\times \sin\phi_0 \sin 2\psi,$$
$$\mathcal{A}^2 \equiv A_+ \cos\phi_0 \sin 2\psi + A_\times \sin\phi_0 \cos 2\psi,$$
$$\mathcal{A}^3 \equiv -A_+ \sin\phi_0 \cos 2\psi - A_\times \cos\phi_0 \sin 2\psi,$$
$$\mathcal{A}^4 \equiv -A_+ \sin\phi_0 \sin 2\psi + A_\times \cos\phi_0 \cos 2\psi, \quad (4)$$

such that the signal $s^X(t)$ of Eq. (1) at a detector $X$ can now be written in the form

$$s^X(t;\mathcal{A},\lambda) = \sum_{\mu=1}^{4} \mathcal{A}^\mu h_\mu^X(t;\lambda), \quad (5)$$

where the detector-dependent basis functions $h_\mu^X(t;\lambda)$ are given by

$$h_1^X(t) \equiv a^X(t) \cos\phi^X(t),$$
$$h_2^X(t) \equiv b^X(t) \cos\phi^X(t),$$
$$h_3^X(t) \equiv a^X(t) \sin\phi^X(t),$$
$$h_4^X(t) \equiv b^X(t) \sin\phi^X(t), \quad (6)$$

in terms of the signal phase $\phi^X(t)$ at detector $X$ and antenna-pattern functions $a^X(t)$ and $b^X(t)$, for which explicit expressions can be found, again, in [28].

## III. THE CW LIKELIHOOD FUNCTION

### A. The $\mathcal{F}$-statistic formalism

The $\mathcal{F}$-statistic is a partially maximized [28] (or marginalized [29]) likelihood ratio between two hypotheses, namely a signal ($\mathcal{H}_\mathrm{S}$) and a noise hypothesis ($\mathcal{H}_\mathrm{N}$). The signal hypothesis $\mathcal{H}_\mathrm{S}$ states that the strain data $x^X(t)$ in detector $X$ contains a signal $s^X(t)$ described by Eq. (5) in addition to (Gaussian) noise $n^X(t)$, namely,

$$x^X(t) = n^X(t) + s^X(t;\mathcal{A},\lambda). \quad (7)$$

The noise hypothesis $\mathcal{H}_\mathrm{N}$, on the other hand, assumes that the data contains only (Gaussian) noise $n^X(t)$, i.e., $s^X = 0$.

For ease of notation we define a multidetector scalar product [30,31] between time series $x^X(t)$ and $y^X(t)$ as

$$(x|y) \equiv 2 \sum_{X}^{N_\mathrm{Det}} S_X^{-1} \int_0^T x^X(t)y^X(t)dt, \quad (8)$$

where $N_\mathrm{Det}$ is the number of detectors and $S_X$ is the (single-sided) noise power spectral density (PSD) of detector $X$ around the narrow frequency band of interest. For simplicity this expression assumes fully stationary noise, but it can be easily generalized [31] to the weaker assumption of stationarity over short time stretches $T_{\text{SFT}}$, which is used in the actual implementation.

With this definition of scalar product, it can be shown [32] that the likelihood function for the Gaussian-noise hypothesis $\mathcal{H}_{\text{N}}$ can be written as

$$P(x|\mathcal{H}_{\text{N}}) = \kappa e^{-\frac{1}{2}(x|x)}, \quad (9)$$

where $\kappa$ is a normalization factor. From Eq. (7) we can therefore also express the likelihood for the signal hypothesis $\mathcal{H}_{\text{S}}$ for a particular signal $s(t; \mathcal{A}, \lambda)$ as

$$P(x|\mathcal{H}_{\text{S}}, \mathcal{A}, \lambda) = \kappa e^{-\frac{1}{2}(x-s|x-s)}. \quad (10)$$

For the detection problem of deciding whether the signal or noise hypothesis is favored by the data $x$, both the frequentist as well as the Bayesian framework require expressing the *likelihood ratio* $\mathcal{L}$ between the two hypotheses, namely,

$$\mathcal{L}(x; \mathcal{A}, \lambda) \equiv \frac{P(x|\mathcal{H}_{\text{S}}, \mathcal{A}, \lambda)}{P(x|\mathcal{H}_{\text{N}})} = e^{(x|s) - \frac{1}{2}(s|s)}, \quad (11)$$

and substituting Eq. (5) for the signal $s$ we can further write this as

$$\log \mathcal{L}(x; \mathcal{A}, \lambda) = \mathcal{A}^{\mu} x_{\mu} - \frac{1}{2} \mathcal{A}^{\mu} \mathcal{M}_{\mu\nu} \mathcal{A}^{\nu}, \quad (12)$$

with implicit summation over repeated amplitude indices $\mu, \nu = 1\ldots4$, and the definitions

$$x_{\mu}(\lambda) \equiv (x|h_{\mu}), \quad \text{and} \quad \mathcal{M}_{\mu\nu}(\lambda) \equiv (h_{\mu}|h_{\nu}). \quad (13)$$

The four numbers $x_{\mu}$ are the "matched filter" scalar products of the data $x$ with the four CW basis functions $h_{\mu}$ of Eq. (6). The symmetric $4 \times 4$ matrix $\mathcal{M}_{\mu\nu}$, often referred to as the *antenna-pattern matrix*, quantifies the response of the detector network for a particular sky direction. For ground-based detectors (using the long-wavelength approximation) the antenna-pattern matrix can be found more explicitly [28] as

$$\mathcal{M}_{\mu\nu} = \gamma \begin{pmatrix} A & C & 0 & 0 \\ C & B & 0 & 0 \\ 0 & 0 & A & C \\ 0 & 0 & C & B \end{pmatrix}, \quad (14)$$

in terms of the coefficients

$$A \equiv \langle a^2 \rangle, \quad B \equiv \langle b^2 \rangle, \quad C \equiv \langle a^* b \rangle, \quad (15)$$

where $\langle . \rangle$ indicates (noise-weighted) time-averaging. The prefactor $\gamma$ is

$$\gamma \equiv \mathcal{S}^{-1} T_{\text{data}}, \quad (16)$$

which characterizes the amount and noise-level of the data, in terms of the overall noise floor $\mathcal{S}$, given by the harmonic mean

$$\mathcal{S}^{-1} \equiv \frac{1}{N_{\text{Det}}} \sum_X S_X^{-1}, \quad (17)$$

and the total amount of data from all detectors, $T_{\text{data}}$. We discuss some of the statistical properties of the log-likelihood ratio in Appendix A.

In practice the implementation uses detector strain data in the form of short Fourier transforms (SFTs) over time spans $T_{\text{SFT}}$, and stationarity of the noise is only assumed over these short spans; see [31]. For a total number $N_{\text{SFT}}$ of input SFTs used from all detectors, the total amount of data is $T_{\text{data}} \equiv N_{\text{SFT}} T_{\text{SFT}}$.

As first shown in [28], using the reparametrization of Eq. (5), the log-likelihood ratio equation (12) is a quadratic function over the $\mathcal{A}^{\mu}$ and can therefore be maximized analytically:

$$\mathcal{F}(x, \lambda) \equiv \max_{\mathcal{A}} \log \mathcal{L}(x; \mathcal{A}, \lambda) = \frac{1}{2} x_{\mu} \mathcal{M}^{\mu\nu} x_{\nu}, \quad (18)$$

where we defined $\mathcal{M}^{\mu\nu}$ as the inverse of the antenna-pattern matrix $\mathcal{M}_{\mu\nu}$ of Eq. (14). The same expression can also be obtained as a partial Bayes factor by marginalizing the likelihood ratio over $\mathcal{A}^{\mu}$ for a specific (albeit unphysical) choice of priors on the $\mathcal{A}$, as shown in [29].

Following the standard $\mathcal{F}$-statistic notation of [28], we introduce the two complex quantities

$$F_{\text{a}}(\lambda) \equiv \frac{1}{\sqrt{2\gamma}} (x_1 - ix_3),$$
$$F_{\text{b}}(\lambda) \equiv \frac{1}{\sqrt{2\gamma}} (x_2 - ix_4), \quad (19)$$

and combining this with the explicit antenna-pattern matrix of Eq. (14), we can obtain the $\mathcal{F}$-statistic in the form

$$2\mathcal{F} = \frac{2}{D} [B|F_{\text{a}}|^2 + A|F_{\text{b}}|^2 - 2C\Re(F_{\text{a}}^* F_{\text{b}})], \quad (20)$$

where $D \equiv AB - C^2$ is the determinant of the nonzero $2 \times 2$ block in $\mathcal{M}_{\mu\nu}$, and $\Re$ denotes the real part.

### B. $\mathcal{F}$-statistic-based likelihood

The $\mathcal{F}$-statistic implementation in LALSuite [33] proceeds by first computing the two complex numbers $F_{\text{a}}(\lambda), F_{\text{b}}(\lambda)$ and the antenna-pattern matrix coefficients $A, B, C$, and then combining them via Eq. (20). However, we see from Eq. (12) that these are the same ingredients needed to express the full likelihood ratio. Specifically, we can express the two terms in the likelihood as

$$\mathcal{A}^\mu x_\mu(\lambda) = \sqrt{2\gamma}(\mathcal{A}^1 F_a^\Re + \mathcal{A}^2 F_b^\Re - \mathcal{A}^3 F_a^\Im - \mathcal{A}^4 F_b^\Im), \quad (21)$$

using real $\Re$ and imaginary $\Im$ parts of the $F_a, F_b$, and

$$\mathcal{A}^\mu \mathcal{M}_{\mu\nu}(\lambda) \mathcal{A}^\nu = h_0^2 \gamma (\alpha_1 A + \alpha_2 B + 2\alpha_3 C)$$
$$\equiv \rho^2(\mathcal{A}, \lambda), \quad (22)$$

which defines the *signal power* $\rho^2$, also known as the squared (perfect-match) signal-to-noise ratio (SNR), and with amplitude angle factors $\alpha_i(\cos \iota, \psi)$

$$\alpha_1 \equiv \frac{1}{4}(1 + \cos^2 \iota)^2 \cos^2 2\psi + \cos^2 \iota \sin^2 2\psi,$$

$$\alpha_2 \equiv \frac{1}{4}(1 + \cos^2 \iota)^2 \sin^2 2\psi + \cos^2 \iota \cos^2 2\psi,$$

$$\alpha_3 \equiv \frac{1}{4}(1 - \cos^2 \iota)^2 \sin 2\psi \cos 2\psi. \quad (23)$$

We can use Eq. (11) to express the signal likelihood function as

$$P(x|\mathcal{H}_S, \mathcal{A}, \lambda) = \mathcal{L}(x; \mathcal{A}, \lambda) P(x|\mathcal{H}_N), \quad (24)$$

where $\mathcal{L}$ can be computed from the byproducts of the $\mathcal{F}$-statistic calculation, namely Eqs. (21) and (22), and the noise likelihood does not depend on any signal parameters.

Note that this likelihood, although derived from an $\mathcal{F}$-statistic framework, is fundamentally the same as that used in the *time domain method* of [20], and is mathematically equivalent to the (non-noise-marginalized) expression in Eq. (19) of [34].

### C. Bayesian parameter-estimation framework

#### 1. Likelihood

In a targeted CW search, the phase-evolution parameters $\lambda$ are assumed to be known from electromagnetic observations, while the amplitude parameters are generally unknown. Using Bayes' theorem, the posterior for the unknown amplitude parameters $\mathcal{A}$ is

$$P(\mathcal{A}|x, \mathcal{H}_S, \lambda) = P(\mathcal{A}|\mathcal{H}_S, \lambda) \frac{P(x|\mathcal{H}_S, \mathcal{A}, \lambda)}{P(x|\mathcal{H}_S, \lambda)}, \quad (25)$$

where $P(\mathcal{A}|\mathcal{H}_S, \lambda)$ is the prior on the amplitude parameters, $P(x|\mathcal{H}_S, \mathcal{A}, \lambda)$ is the signal likelihood derived in Sec. III B, and $P(x|\mathcal{H}_S, \lambda)$ is the amplitude-marginalized signal likelihood. Using Eq. (24) and collecting all $\mathcal{A}$-independent factors into a proportionality constant $k$, this yields

$$P(\mathcal{A}|x, \mathcal{H}_S, \lambda) = k\mathcal{L}(x; \mathcal{A}, \lambda) P(\mathcal{A}|\mathcal{H}_S, \lambda), \quad (26)$$

where $k$ can be determined via the normalization condition $\int P(\mathcal{A}|\ldots) d^4\mathcal{A} = 1$.

#### 2. Priors

Typically we have weak or no prior information on the intrinsic amplitude of the signal $h_0$ and the angle parameters $\{\iota, \psi, \phi_0\}$ of the source.

If there are no observational constraints on the rotation axis of the pulsar, we assume isotropic "ignorance" priors on the angle parameters, following the standard choices that we recap below [26,29,34]:

(i) The initial phase $\phi_0$ corresponds to the pulsar rotation angle at a reference time and the ignorance prior is uniform over the range $\phi_0 \in [0, 2\pi)$.
(ii) The ignorance prior for the direction of the rotation axis is also uniform $\in [0, 2\pi)$ and it translates to uniform priors in $\cos \iota \in [-1, 1]$ and $\psi \in [0, 2\pi)$.
(iii) From Eq. (4) we see that $\psi \to \psi + \pi$ leaves the $\mathcal{A}^\mu$ unchanged, and further that $\psi \to \psi + \pi/2$ flips their sign, which can be compensated by $\phi_0 \to \phi_0 + \pi$. We can therefore choose a gauge where $\psi \in [-\pi/4, \pi/4]$ and $\phi_0 \in [0, 2\pi)$.

When pulsar observations do constrain these priors, they can be modified appropriately.

When it comes to $h_0$, the choice of prior range $[h_{\text{low}}, h_{\text{high}}]$ and probability distribution is less straightforward and ultimately depends on the specific case being considered.

When targeting a known pulsar, one could inform the $h_{\text{high}}$ from the observed pulsar parameters, namely, the spin-down upper limit $h_0^{\text{sd}}$ of Eq. (34), which indicates the maximal possible amplitude of a CW signal if all the rotational energy lost by the pulsar was converted into gravitational waves. One could, therefore, require $h_{\text{high}} \leq h_0^{\text{sd}}$.

If a previous targeted search has established an $h_0$ upper limit $h_0^{\text{UL}}$ for the pulsar, then, under the assumption that the signal amplitude does not change over time, one could require $h_{\text{high}} < h_0^{\text{UL}}$.

Another possibility is to use physical estimates on the possible range of ellipticities of neutron stars, which is a measure of the nonaxisymmetric deformation defined as

$$\varepsilon = \frac{|I_{xx} - I_{yy}|}{I_{zz}}, \quad (27)$$

where $I_{aa}$ denotes the moment of inertia of the object along axis $a$. One can then derive an $h_0$ prior range from the range of possible $\varepsilon$ using (e.g., see [35])

$$h_0(\varepsilon) = \frac{4\pi^2 G \varepsilon I_{zz} f^2}{c^4 d}, \tag{28}$$

where $f$ is the CW signal frequency and $d$ is the distance to the pulsar. The maximum deformation $\varepsilon^{\max}$ that the neutron star crust can sustain before breaking also provides an indication of the largest possible gravitational wave amplitude.

Putting all these considerations together one could argue that $h_{\text{high}}$ should be the smallest among (i) the spin-down limit $h_0^{\text{sd}}$, (ii) the amplitude corresponding to the largest sustainable deformation, and (iii) the largest amplitude compatible with previous observations.

Quadrupolar deformations can also be sourced by an internal magnetic field $B$, and are predicted to be very small. Usually, the smallest signal amplitudes correspond to this sort of mechanism, so one could place $h_{\text{low}} \sim h_0(B)$. We refer the reader to [36] and references therein for further discussion.

When a strong signal is present, the prior has minimal influence on the resulting posterior, because the likelihood will be strongly peaked. In the realm of a weak or nondetectable signal, however, a uniform prior on $h_0$ leads to a more "conservative" (i.e., higher) upper limit compared to a log-uniform distribution, as discussed in [37]. A prior distribution uniform in the logarithm, on the other hand, ensures a uniform sampling when our ignorance spans several orders of magnitude. To alleviate the concern that an upper limit based on log-uniform $h_0$ priors is range dependent, [37] showed that such dependence is, in fact, weak.

### 3. Software

We use the "Demod" [38] implementation of the $\mathcal{F}$-statistic within the LALSuite [33] software library for the $F_{\{a,b\}}$ calculation. This uses Fourier transforms of the data—the SFTs—computed over relatively short periods of time, such that the instantaneous signal frequency does not move during that time period by more than a Fourier bin. The method involves the usage of the Dirichlet kernel [31], which peaks at the frequency ($k^*$) of the signal on the SFT data. For efficiency, the kernel is approximated by truncating it to a few bins ($\Delta k$) on each side of $k^*$ for computational efficiency. We use a $\Delta k$ of 8 bins unless stated otherwise.

The signal likelihood function expressed in Sec. III B is used with a stochastic sampler to compute the likelihood in the $\{h_0, \cos\iota, \psi, \phi_0\}$ parameter space weighted by the prior. See [39] for a discussion on stochastic sampling. For this, we use the Python library BILBY [40], specifically the class *core.sampler*, to interface the different available Python samplers with our likelihood function. Since the latter is implemented in C99, we use the SWIGLAL wrapper of [41] to pass it to BILBY in Python.

### 4. Timing

On an Intel(R) Xeon(R) CPU E5-2620 v4 @ 2.10 GHz processor, the median time for a single computation of the likelihood function is $\sim \mathcal{O}(\mu s)$ with the $\phi_0$-marginalized likelihood of Sec. IV needing around twice as much time than the full likelihood. This processor is slower than the one used by [20], where it was reported that their standard likelihood takes 81 μs to be computed.

The full runtime of the pipeline depends on the size of the prior space to probe and the number of CPU cores that can be used in parallel. For the known-pulsar search reported in Sec. VI, with 32-core parallelization, the search in the four-dimensional parameter space took 154 seconds whereas the search in the three-dimensional space (using the $\phi_0$-marginalized likelihood of Sec. IV) concluded in 124 seconds. Note that we reused the narrow-banded SFTs already prepared for the templated search that was reported in [18], so the cost of data-preparation steps prior to the actual search is not included here.

### D. Tests

#### 1. Recovery of a simulated signal

The first test of the method is to check if it correctly recovers the parameters of a simulated signal. We test in the absence of noise to avoid the signal peak in the posterior getting shifted from the injection point.

We simulate a one-year-long signal of $h_0 \approx 4 \times 10^{-27}$ in the H1 and L1 detectors and compute the likelihood assuming a noise floor of $1 \times 10^{-25}/\sqrt{\text{Hz}}$, corresponding to an SNR of $\rho \approx 100$. The phase parameters of the signal are given in Table I. The year-long data is converted into SFTs of 10 s in this test. The nested sampler DYNESTY [42] is used with options $nlive = 5000$ and $dlogz = 0.01$. The priors for $\{\cos\iota, \psi, \phi_0\}$ are as described in Sec. III C 2. For this test, we choose a simple uniform $h_0$ prior in the range $[10^{-28}, 7.1 \times 10^{-27}]$ centered on the $h_0$ of the injected signal.

TABLE I. Settings of injection and recovery tests in Figs. 1, 2, 5, and 6: SFT time base, number of Dirichlet-kernel bins, sampler options, and phase-evolution parameters of a fake signal. True values of amplitude parameters are shown in the figures as orange lines.

| Parameter | Value |
| --- | --- |
| $T_{\text{SFT}}$ (seconds) | 10 |
| $\Delta k$ (bins) | 8 |
| $nlive$ | 5000 |
| $dlogz$ | 0.01 |
| Start of fake signal (global positioning system (GPS)) | 1234567890.0 |
| Reference epoch (GPS) | 1242451890.0 |
| Right Ascension, $\alpha$ (rad) | 0.26 |
| Declination, $\delta$ (rad) | 0.30 |
| CW Frequency, $f$ (Hz) | 100.0 |
| CW Frequency Derivative, $\dot{f}$ (Hz s$^{-1}$) | $-1 \times 10^{-15}$ |

Figure 1 shows the posterior distributions recovered by our pipeline. The true amplitude parameters of the injected signal, indicated by the orange lines, are accurately recovered by the method at the maximum of the posterior, as is expected in the absence of noise.

As discussed in Sec. III C 2, the signal is degenerate under the transformation $\psi \to \psi + \pi/2, \phi_0 \to \phi_0 + \pi$. This results in a bimodal posterior distribution in the $\psi - \phi_0$ subspace when the $\psi$ value of the signal is close to the edges of its $[-\pi/4, \pi/4)$ range. A nested sampler like DYNESTY tends to handle multimodal likelihoods better than an MCMC sampler (see, e.g., [43,44]). In Fig. 2, the simulated signal of Table I but with $\psi = -\pi/4$ is recovered using DYNESTY, with two modes in the posterior split by $\pi/2$ radians in $\psi$ and by $\pi$ radians in $\phi_0$. In Fig. 3, the same signal is searched using BilbyMCMC [45] with the default parameters. It recovers only one of the modes in the $\psi - \phi_0$ subspace. Owing to its better performance in multimodal parameter spaces, DYNESTY is chosen as the default sampler for the rest of this paper.

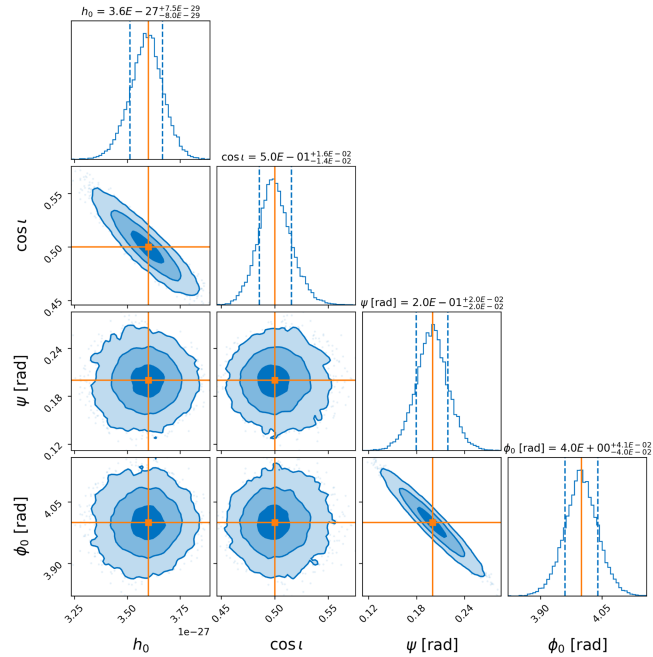

FIG. 1. Corner plot showing the recovery of a simulated signal ($\rho \approx 100$) without noise. The true values of the signal amplitude parameters are shown in orange. The blue vertical lines show the 16th and 84th percentile of the distribution, and together they bracket a 68% credible interval in the high probability-density region. The title for each 1D posterior plot shows the median value and the 1-$\sigma$ error of the parameter. The 2D isoprobability levels contain $\approx 39\%$, $\approx 87\%$ and $\approx 99\%$ of the posterior area corresponding to the 1-,2- and 3-$\sigma$ levels of a two-dimensional Gaussian distribution.

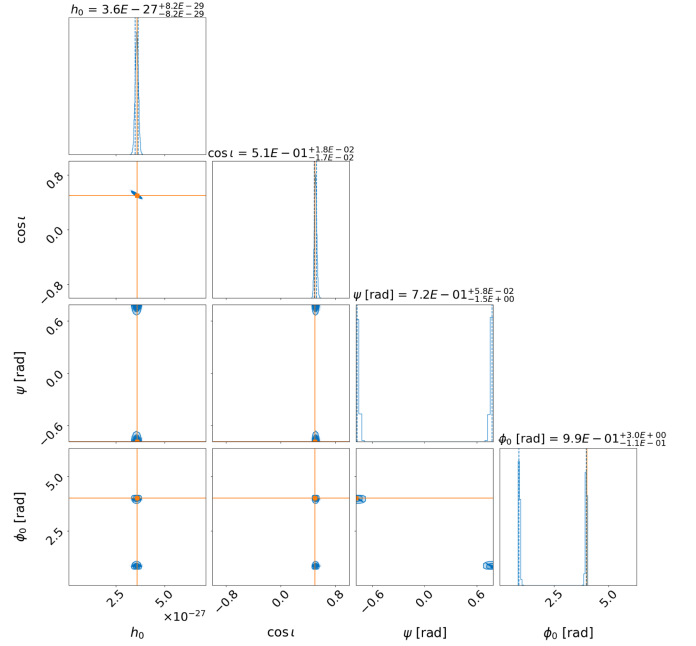

FIG. 2. The fake signal of Fig. 1 with the modification of $\psi = -\pi/4$ as recovered by the DYNESTY sampler. The multimodal posteriors in $\psi - \phi_0$ parameter space are recovered by this nested sampler.

### 2. Percentile-percentile plots

A second test is the percentile-percentile (PP) plot, which checks whether the Bayesian credible intervals on the posterior distributions of parameters, as returned by the method, correspond to frequentist confidence intervals when sampling from the priors. To do this, we construct 10 000 fake signals whose amplitude parameters are drawn randomly from their priors and whose phase-evolution parameters are fixed to the values in Table II. The $h_0$ prior is log uniform in the range $[1 \times 10^{-28}, 4 \times 10^{-26}]$. We embed the signals in simulated Gaussian noise contiguously spanning the three observation runs from the two Advanced LIGO detectors (henceforth O1O2O3) with a noise floor of $9 \times 10^{-24}/\sqrt{\text{Hz}}$. The SFT time baseline is $T_{\text{SFT}} = 1800$ s. The highest SNR of a simulated signal in this test is $\rho \approx 50$. We use DYNESTY with $nlive = 500$ and $dlogz = 0.1$, a less stringent convergence criterion than used in the previous section to reduce the computational cost of this test.

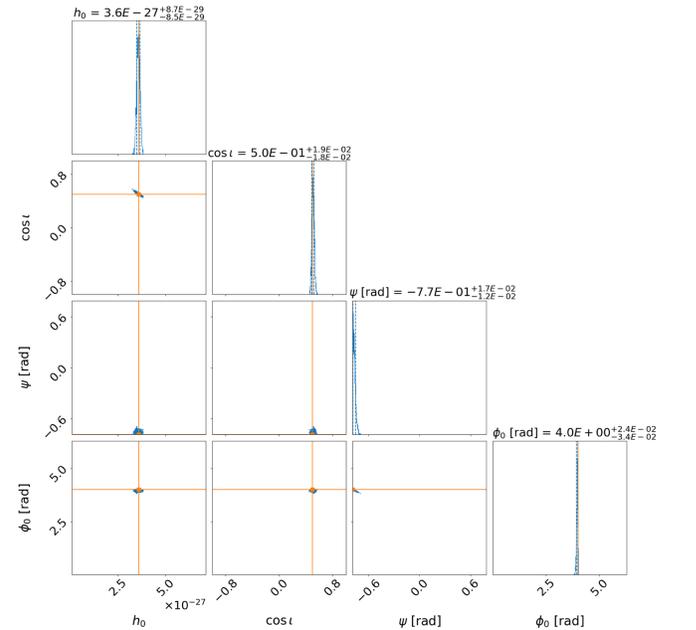

FIG. 3. The fake signal of Fig. 1 with the modification of $\psi = -\pi/4$ as recovered by the BilbyMCMC sampler, with $nsamples = 1000$ under its default settings. The sampler recovers only one mode of the posterior in $\psi - \phi_0$ space.

TABLE II. Settings of PP plot tests in Figs. 4, 7, and 14: Time span of data ($T_{\text{span}}$), Time base of SFTs ($T_{\text{SFT}}$), number of bins in Dirichlet kernel ($\Delta k$), sampler options, and phase-evolution parameters of fake signals.

| Parameter | Value |
|---|---|
| $T_{\text{span}}$ (seconds) | 142739988.0 |
| $T_{\text{SFT}}$ (seconds) | 1800 |
| $\Delta k$ (bins) | 8 |
| Sampler | dynesty |
| $nlive$ | 500 |
| $dlogz$ | 0.1 |
| Start of fake signals (GPS) | 1126623625.0 |
| Reference epoch (GPS) | 1081123148.8 |
| Right Ascension, $\alpha$ (rad) | 1.13 |
| Declination, $\delta$ (rad) | 1.16 |
| CW Frequency, $f$ (Hz) | 687.24 |
| CW Frequency Derivative, $\dot{f}$ (Hz s$^{-1}$) | $-3.2 \times 10^{-15}$ |

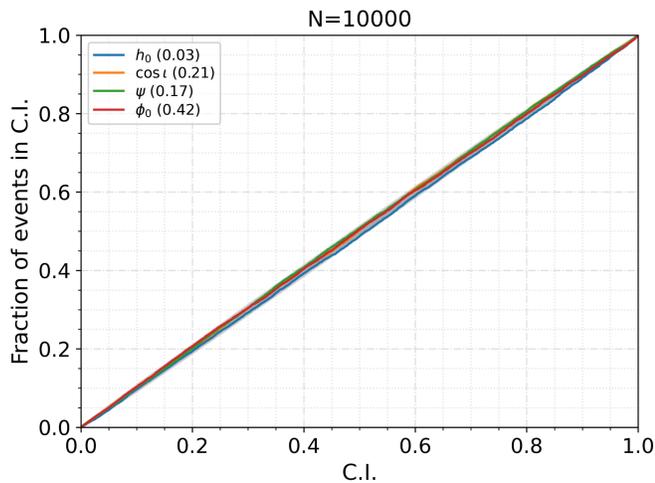

FIG. 4. PP plot with maximum $\rho \approx 50$. We sample 10 000 injections from the prior range $h_0 \in [10^{-28}, 4 \times 10^{-26}]$ and the full ranges as mentioned in Sec. III C 2 for the other amplitude parameters. The legend shows the per-parameter KS $p$-values.

Ideally, $x\%$ of the total number of injections should fall in the $x\%$ credible interval. This corresponds to a uniform distribution of the measured credible intervals. We test that this is the case with a Kolmogorov-Smirnov (KS) test, quantifying the conformity of the two distributions with a $p$-value (higher $p$-values imply better agreement). The results are shown in Fig. 4.

Although the KS $p$-value shown in Fig. 4 for $h_0$ is quite small and indicates some level of systematic bias (which we describe in Appendix B), we argue that in practice this does not pose a critical issue. As can be seen in the figure, although the $h_0$ KS $p$-value is small, the absolute error in the percentage of recovered signals is actually quite small. For example, 89.2% of the signals fall within the 90% credible interval for $h_0$.

## IV. $\phi_0$-MARGINALIZED LIKELIHOOD

### A. Expression for the marginalized likelihood $\mathcal{L}^{\overline{\phi_0}}$

As shown in [46] (Sec. 5.4), the likelihood ratio $\mathcal{L}$ of Eq. (24) can be analytically marginalized over $\phi_0$.

This has several advantages for parameter estimation: it avoids the bimodality of posteriors in $\psi$-$\phi_0$ discussed in the previous section, and it leaves us with fewer dimensions to explore numerically. For example, for the purpose of calculating $h_0$ upper limits, this tends to yield better numerical robustness and accuracy. Additionally, the $\phi_0$-marginalized likelihood provides a consistency check for the results from full-likelihood.

From Eq. (4) we can explicitly factor out the $\phi_0$ dependence in the $\mathcal{A}^\mu x_\mu$ term that appears in the likelihood-ratio equation (24):

$$\mathcal{A}^\mu x_\mu = q_s \sin \phi_0 + q_c \cos \phi_0$$
$$= q \cos(\phi_0 - \varphi_0), \quad (29)$$

with

$$q_s \equiv -\sin 2\psi(x_1 A_\times + x_4 A_+) + \cos 2\psi(x_2 A_\times - x_3 A_+),$$
$$q_c \equiv \cos 2\psi(x_1 A_+ + x_4 A_\times) + \sin 2\psi(x_2 A_+ - x_3 A_\times), \quad (30)$$

and $\tan \varphi_0 \equiv q_s/q_c$, and

$$q^2(x; h_0, \cos\iota, \psi) \equiv q_s^2 + q_c^2$$
$$= 2h_0^2 \gamma[\alpha_1 |F_a|^2 + \alpha_2 |F_b|^2 + 2\alpha_3 \Re(F_a^* F_b)]. \quad (31)$$

We can see from Eq. (22) that the signal power $\rho^2$ does not depend on $\phi_0$, and therefore writing the likelihood ratio in the form

$$\mathcal{L}(x; \mathcal{A}) = e^{-\frac{1}{2}\rho^2} e^{q\cos(\phi_0 - \varphi_0)}, \quad (32)$$

makes the $\phi_0$ dependence fully explicit. Using the uniform $\phi_0$ prior of Sec. III C 2, we can now obtain the $\phi_0$-marginalized likelihood ratio $\mathcal{L}^{\overline{\phi_0}}$ in the form

$$\begin{aligned}\mathcal{L}^{\overline{\phi_0}}(x; h_0, \cos\iota, \psi) &\equiv \int_0^{2\pi} \mathcal{L}(x; \mathcal{A}) P(\phi_0|\mathcal{H}_S) d\phi_0 \\ &= \frac{1}{2\pi} \int_0^{2\pi} \mathcal{L}(x; \mathcal{A}) d\phi_0 \\ &= \frac{1}{2\pi} e^{-\frac{1}{2}\rho^2} \int_0^{2\pi} e^{q \cos(\phi_0 - \varphi_0)} d\phi_0 \\ &= e^{-\frac{1}{2}\rho^2} I_0(q), \end{aligned} \quad (33)$$

where we used the Jacobi-Anger expansion [47] to see that $\int_0^{2\pi} e^{q \cos\phi} d\phi = 2\pi I_0(q)$, in terms of the modified Bessel function of the first kind $I_0$.

### B. Tests

#### 1. Recovery of a simulated signal

We test again the recovery of the simulated signal (without noise) of Sec. III D 1, this time using the $\phi_0$-marginalized likelihood $\mathcal{L}^{\overline{\phi_0}}$. The resulting posterior distributions on $\{h_0, \cos\iota, \psi\}$ are shown in blue in Fig. 5, indicating that these parameters have again been accurately recovered by the method, as they coincide with the maximum of the posterior.

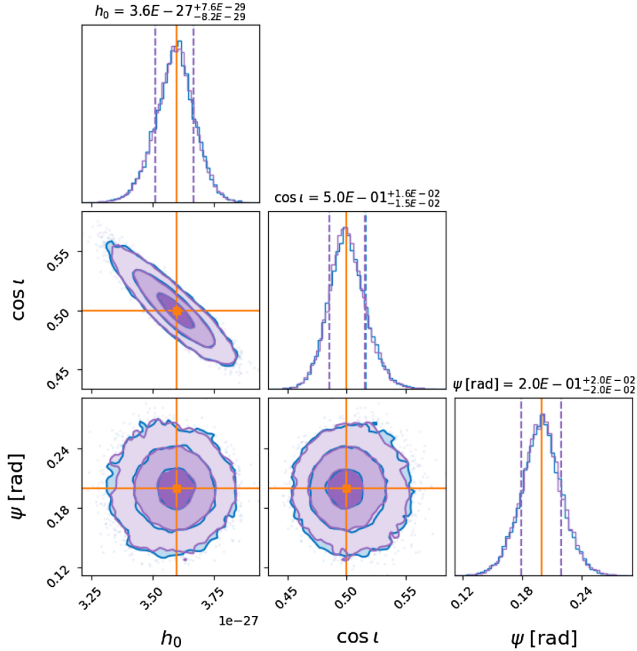

FIG. 5. Blue: The posterior distributions on $\{h_0, \cos\iota, \psi\}$ in the absence of noise for the signal of Sec. III D 1, analyzed using the $\phi_0$-marginalized likelihood. See caption of Fig 1 for description. Purple: The posteriors from the full likelihood (same as in Fig. 1) are overlaid on the posteriors from the $\phi_0$-marginalized likelihood to show that the analytical marginalization (blue) and the numerical marginalization over $\phi_0$ agree with each other. The true injection parameters are shown in orange.

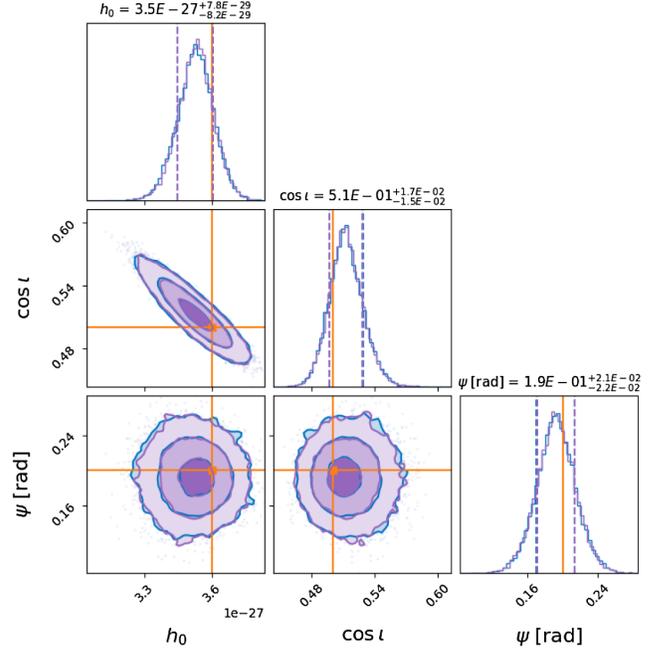

FIG. 6. Recovery of the simulated signal of Sec. III D 1 by the $\phi_0$-marginalized likelihood (in blue) and the full likelihood (in purple) in the presence of noise. The true parameters of the signal are shown in orange. The posteriors peak away from these true values because of the presence of noise.

Additionally, the posteriors on $\{h_0, \cos\iota, \psi\}$ computed by the four-dimensional likelihood of Sec. III and numerically marginalized over $\phi_0$, and by the three-dimensional ($\phi_0$-marginalized) likelihood should be equivalent. To show that this is indeed the case in the noiseless scenario, the posteriors from the full likelihood are overlaid in purple on the posteriors from the $\phi_0$-marginalized likelihood in Fig. 5.

To test that this is true also when noise is present, we search for the simulated signal of Sec. III D 1 with Gaussian noise (with a noise floor of $10^{-25}/\sqrt{\text{Hz}}$) using the two likelihood functions. A comparison of the resulting posteriors is shown in Fig. 6. The $\phi_0$-marginalized likelihood (in blue) and the full likelihood (in purple) produce posteriors that are consistent with each other. The peaks of both sets of posteriors deviate from the true values, as is expected in the presence of noise.

#### 2. PP plots

Next, we produce PP plots for the $\phi_0$-marginalized likelihood, as was done for the full likelihood in Sec. III D 2. We simulate signals with amplitudes from the range $h_0 \in [10^{-28}, 4 \times 10^{-26}]$ in contiguous O1O2O3 data, with a maximum signal SNR of $\rho \approx 50$, with the same setup of Table II. The results with the per-parameter KS-test

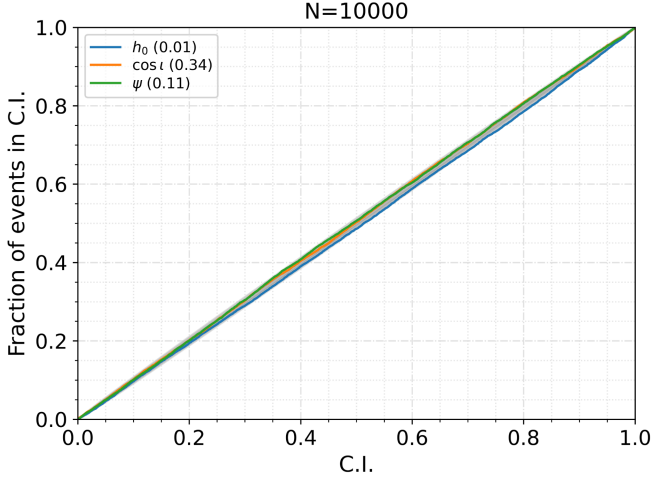

FIG. 7. PP plot using the $\phi_0$-marginalized likelihood with maximum $\rho \approx 50$.

$p$-values are shown in Fig. 7. The KS $p$-values are of similar magnitude to the ones in Fig. 4.

However, a PP test with (unrealistically) high SNR signals in the range of up to $\rho \approx 1000$ reveals an increasing bias in the $h_0$ posterior, indicating a current limitation of the method, affecting both the full as well as the marginalized-$\phi_0$ likelihoods. We discuss this problem and its underlying causes in Appendix B. However, such high signal strengths are unrealistic in the present-day scenario of ground-based continuous gravitational wave searches, and solving this issue is beyond the scope of this paper.

## V. RECOVERY OF HARDWARE INJECTIONS

We apply the parameter-estimation method on the CW hardware injections present in the data of the Advanced LIGO detectors. Namely, we search for 17 of the 18 hardware injections[1] added in O3a data. We use both the full likelihood of Sec. III as well as the $\phi_0$-marginalized likelihood of Sec. IV, but for simplicity we present here only the results from the latter. The phase-evolution parameters of each search are fixed at those of the respective hardware injection. We use isotropic priors on $\cos \iota$ and $\psi$ as discussed in Sec. III C 2, and a log-uniform amplitude prior in the range $h_0 \in [10^{-28}, 10^{-23}]$, which includes the true $h_0$ of all hardware injections. The DYNESTY nested sampler is used with $nlive = 5000$ and $dlogz = 0.01$.

Note that here we cannot perform a PP-style consistency test of how many injections are found within which percentiles because the injections were not drawn from a prior that we know. But given the small number of injections, we would not expect to find signals in the tails of the posteriors. Table III shows for the targeted hardware injections, the number of standard deviations ($\sigma$s) in the distance between the maximum posterior point and the true value of each of $\{h_0, \cos \iota, \psi\}$ parameters, along with the SNR of the maximum posterior point. In the case of hardware injections, it can be difficult to identify the cause of larger deviations, given there can be inaccuracies in the actuation forces that generated the hardware injections, as well as non-Gaussian noise artifacts in the data that can affect the results. Therefore we also employ *lalapps_knope* of [20] to recover the hardware injections and cross-check against our results.

[1]"Pulsar 15" at 2991 Hz is omitted for simplicity as there were no SFTs readily available at that high frequency.

TABLE III. Recovery of hardware injections. For every Pulsar we indicate the recovered SNR $\rho$ of the injection and the distance ($\Delta$) between the maximum posterior point and the true value of the parameters $h_0, \cos \iota, \psi$, in terms of standard deviations ($\sigma$s) of their 1D posterior distributions.

| Pulsar ID | SNR ($\rho$) | $\Delta h_0/\sigma_{h_0}$ | $\Delta \cos \iota / \sigma_{\cos \iota}$ | $\Delta \psi / \sigma_\psi$ |
|---|---|---|---|---|
| 0 | 29.9 | 1.35 | 1.42 | 2.83 |
| 1 | 102.1 | 2.48 | 7.25 | 1.70 |
| 2 | 32.6 | 2.67 | 2.57 | 0.93 |
| 3 | 24.8 | 1.89 | 0.76 | 2.35 |
| 4 | 97.3 | 15.49 | 9.28 | 1.70 |
| 5 | 74.3 | 3.10 | 1.50 | 0.89 |
| 6 | 86.1 | 1.30 | 1.01 | 1.51 |
| 7 | 31.1 | 1.90 | 2.77 | 1.76 |
| 8 | 26.0 | 0.87 | 1.47 | 0.95 |
| 9 | 32.9 | 0.15 | 0.38 | 0.15 |
| 10 | 55.6 | 2.48 | 2.65 | 2.40 |
| 11 | 17.2 | 1.77 | 0.41 | 1.21 |
| 12 | 18.4 | 1.08 | 1.00 | 0.86 |
| 13 | 0.4 | 0.25 | 0.13 | 0.24 |
| 14 | 84.8 | 29.99 | 12.05 | 2.88 |
| 16 | 68.2 | 0.28 | 0.11 | 0.83 |
| 17 | 21.1 | 0.49 | 0.19 | 0.33 |

We show the posterior distributions on the amplitude parameters of two hardware injections, Pulsar 3 and 6 in Figs. 8 and 9, respectively.

Below we discuss the hardware injections for which the recovered posteriors are far away from the true $\{h_0, \cos \iota, \psi\}$ values or are noninformative, ordered by how certain we are of what caused the subpar recovery:

*Pulsar 5* lies at 52.8 Hz where PSD plots indicate the presence of non-Gaussian noise artifacts that degrade the recovery.

*Pulsars 14, 4, and 1* at 1991.1, 1390.8, and 848.9 Hz, respectively, are recovered in H1 and L1 separately within $3\sigma$ credible region. The large $\Delta$s in the multidetector search is likely due to an error in the actuation function used for L1 hardware injections, which impairs coherent H1-L1 injection recovery, especially for high-frequency injections (see caption of Table IV in [48]). A discrepancy between the $\phi_0$ values in H1 and L1 is seen by both our method (using full likelihood) and by *lalapps_knope*. Our posteriors on the $h_0$ in single and multidetector searches for these injections are consistent with those from *knope*.

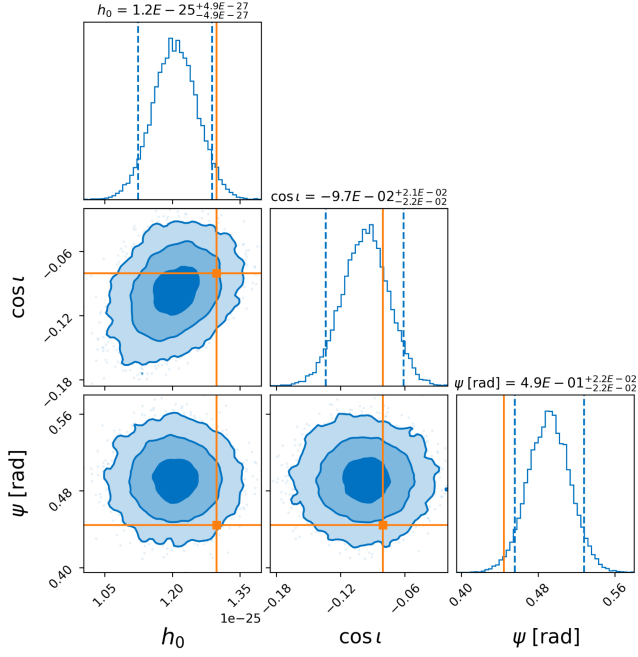

FIG. 8. Posterior probability distribution for the amplitude parameters of the hardware injection "Pulsar 3" in O3a data using the $\phi_0$-marginalized likelihood. The true values of the hardware injection are marked in orange. The dashed lines indicate the 5% and 95% quantiles of the distribution.

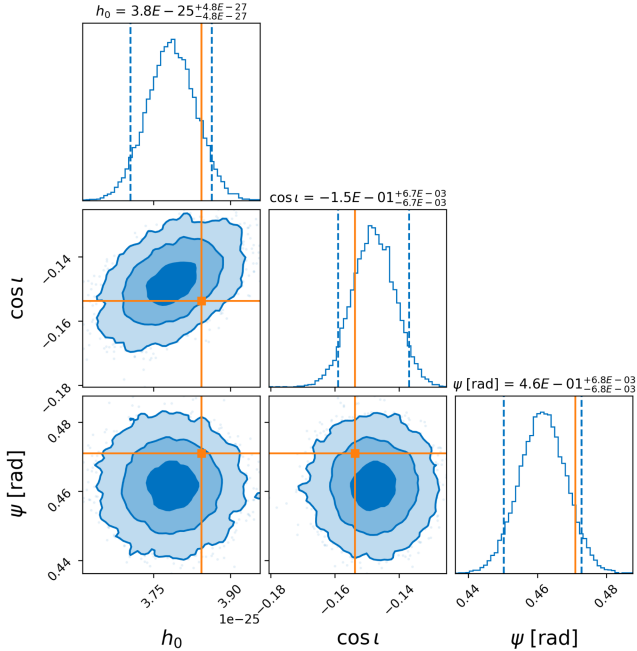

FIG. 9. Posteriors for the amplitude parameters of the hardware injection "Pulsar 6" in O3a data. See Fig. 8 for details.

## VI. FIRST APPLICATION TO SEARCH FOR EMISSION FROM PSR J1526-2744

As a first "real-world" application of the method, we apply it to PSR J1526-2744, which was discovered in a joint survey by TRAPUM and FERMI-LAT [18]. Among the nine pulsars discovered in the survey, PSR J1526-2744 is the only pulsar whose timing could be solved, and the solution is derived from 13 years of FERMI-LAT data that overlap with the Advanced LIGO observation runs.

The pulsar parameters are given in Table IV. PSR J1526-2744 is a binary pulsar at a distance $d$ of 1.3 kpc with spin-down upper limit of

$$h_0^{\rm sd} = \left(\frac{5}{2}\frac{GI_{zz}|\dot{\nu}|}{c^3 d^2 \nu}\right)^{1/2} \approx 7 \times 10^{-28}, \quad (34)$$

where $\nu$ and $\dot{\nu}$ are the pulsar's rotational frequency and spin-down, and $I_{zz}$ is its principal moment of inertia assumed to be the canonical value of $10^{38}$ kg m$^2$.

In [18] we reported single-template and narrow-band continuous wave search results and frequentist upper limits for the emission from the pulsar. Here we describe the Bayesian targeted search for continuous waves from PSR J1526-2744 using our new parameter-estimation pipeline. We assume emission at twice the spin frequency of the pulsar. This is the only mode of emission if the deformed neutron star rotates about one of its principal axis (triaxial aligned model of [49]) and one of the two dominant modes in the more general triaxial nonaligned case of [49].

We use a coherent combination of data from the O1, O2, and O3 observation runs [50] of the Advanced LIGO detectors, gated to remove loud and short glitches in the time domain [51], cleaned to remove narrow lines in the frequency domain, and Fourier transformed with a time base of $T_{\rm SFT} = 60$ s.

The phase-evolution parameters of the search (including the binary orbital parameters) are fixed at the values prescribed by the timing solution from [18]; see Table IV. We use a log-uniform distribution in amplitude in the range $h_0 \in [10^{-28}, 4 \times 10^{-22}]$, based on an ellipticity range of $\varepsilon \in [1.9 \times 10^{-10}, 7.6 \times 10^{-4}]$ for this pulsar, probing below the expected minimum ellipticity of millisecond pulsars [52] and up to (slightly above) the maximum neutron star ellipticity according to [53].

TABLE IV. Gravitational wave parameters of pulsar PSR J1526 − 2744. The uncertainty on the last digit is written inside the parenthesis, as in Table 3 of [18].

| Parameter | Value |
| --- | --- |
| Reference epoch (MJD) | 59355.47 |
| Right Ascension, $\alpha$ | $15^{\rm h}26^{\rm m}45.^{\rm s}103(2)$ |
| Declination, $\delta$ | $-27°44'5.''91(8)$ |
| CW Frequency, $f$ (Hz) | 803.4892041950(5) |
| CW Frequency Derivative, $\dot{f}$ (Hz s$^{-1}$) | $-1.142(2) \times 10^{-15}$ |
| Orbital period, $P$ (days) | 0.2028108285(7) |
| Projected semi-major axis, $a_p$ (lt-s) | 0.22410(3) |
| Epoch of ascending node, $t_{\rm asc}$ (MJD) | 59303.20598(1) |

We perform two searches, one using the full-likelihood function of Sec. III and one with the $\phi_0$-marginalized likelihood function of Sec. IV. We use DYNESTY with options $nlive = 5000$ and $dlogz = 0.01$, which produces a posterior distribution with a total of 8516 samples across the four amplitude parameters. In the case of the marginalized likelihood, using DYNESTY with the same options, the posterior distribution contains 6532 samples over the non-$\phi_0$ amplitude-parameter space.

The resulting posterior distributions on the signal parameters are shown in Figs. 10 and 11. The $h_0$ posterior is consistent with expectations from noise. Since we do not see a signal, the posterior distributions of other signal parameters are noninformative. The 95% upper limit on $h_0$ is obtained by integrating the $h_0$ posterior up to the value of $h_0$ such that 95% of the distribution lies below it. The $h_0^{95\%}$ value for PSR J1526-2744 is found as $6.5 \times 10^{-27}$ from the full likelihood and $6.7 \times 10^{-27}$ from the $\phi_0$-marginalized likelihood. This is a factor of 9.2 larger than the spin-down upper limit of the pulsar and a factor of 1.9 smaller than the frequentist upper limit reported in [18].

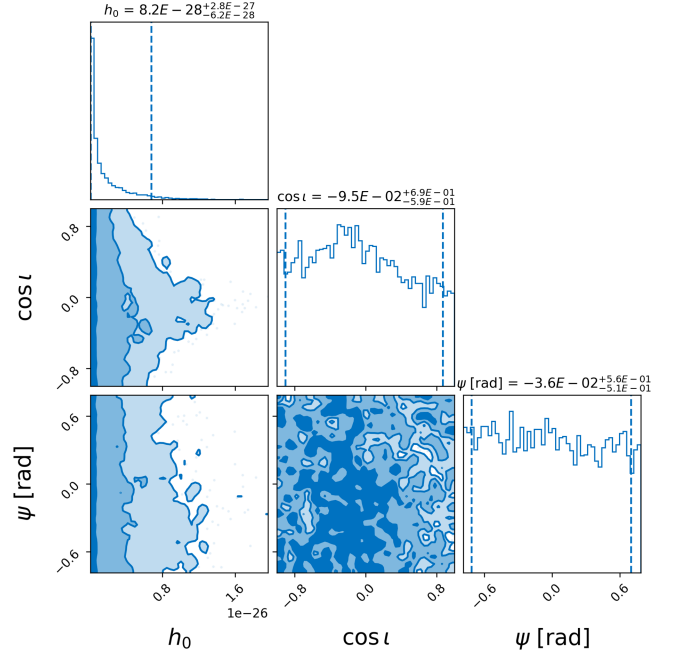

FIG. 11. Bayesian posterior distribution on the non-$\phi_0$ amplitude parameters of the continuous-gravitational-wave signal from PSR J1526-2744 using the $\phi_0$-marginalized likelihood function of Sec. IV. See Fig. 10 for details.

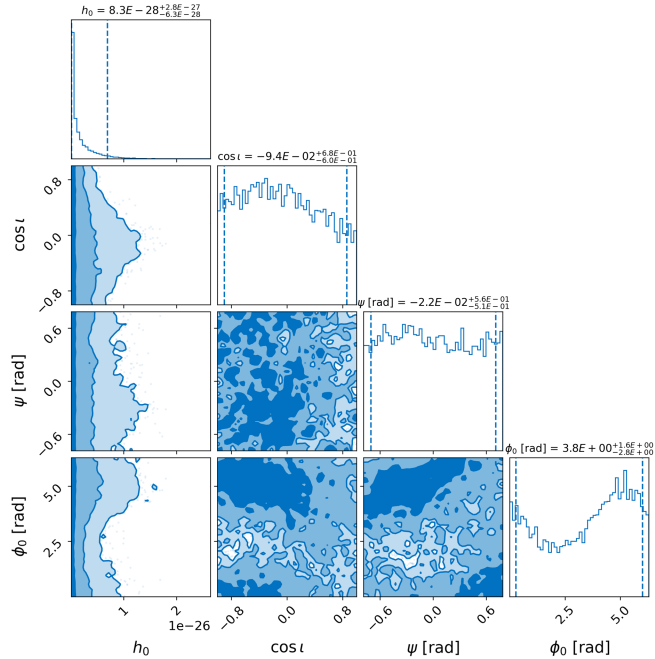

FIG. 10. Bayesian posterior distribution on the amplitude parameters of the continuous-gravitational-wave signal from PSR J1526-2744 using the full-likelihood function of Sec. III. The vertical blue lines show the 5% and 95% quantiles of the distribution, bracketing a 90% credible interval for the parameter. The title for each 1D posterior plot shows the median value and the 1-$\sigma$ error of the parameter.

Bayesian and frequentist upper limits are answers to different questions, and we do not expect them to be numerically the same. The Bayesian asks: given the data and the prior beliefs, what is the value such that with 95% probability, the true value lies below it? In frequentism, 95% is the fraction of repeated trials where in the presence of a signal with parameters at the upper limit value, the procedure recovers the signal. See [54] for a comparison of the two. A factor of $\approx 2$ difference between the two upper limits is not uncommon in known-pulsar searches, as seen, for example, in [26] and Table 4 in [14].

This $h_0^{95\%}$ upper limit can be translated into a 95% confidence upper limit on the ellipticity $\varepsilon$:

$$\varepsilon^{95\%} = 1.3 \times 10^{-8} \left(\frac{h_0^{95\%}}{6.7 \times 10^{-27}}\right)$$
$$\times \left(\frac{d}{1.3 \text{ kpc}}\right) \left(\frac{803.5 \text{ Hz}}{f}\right)^2 \left(\frac{10^{38} \text{ kg m}^2}{I_{zz}}\right). \quad (35)$$

## VII. CONCLUSION AND DISCUSSION

New pulsars are being discovered at a rate faster than ever before and when their timing solution is known they can be targeted for continuous gravitational wave emission with exquisite sensitivity. The most recent constraints are approaching the regime of the expected minimum ellipticity for neutron stars as proposed by [52], making targeted searches ([14,16,55]) a very relevant class of continuous-gravitational-wave searches.

In this paper we introduce and demonstrate a new Bayesian parameter-estimation pipeline, combining well-established machinery from the $\mathcal{F}$-statistic, LALSuite, and BILBY, in order to search for continuous gravitational waves from known pulsars. Previously, only a single Bayesian pipeline existed for such amplitude-parameter estimation [20,56], which operates in the time domain exploiting the knowledge of the signal to reduce the amount of the data to be analyzed via heterodyning, low-pass filtering, and downsampling. Our method works in the frequency domain using only a limited bandwidth of frequency of data decided by the evolution of the signal frequency. At its core, this method is based on the computation of the $\mathcal{F}$-statistic [28,30], utilizing its components to compute the likelihood function.

We use the method to estimate the amplitude parameters of continuous-wave hardware injections in O3a data. Of the 17 hardware injections we targeted, the true $\{h_0, \cos\iota, \psi\}$ of all but 5 are recovered within their $3\sigma$ credible intervals. Of these five, for 1 the posterior remains uninformative. The true parameters of the remaining 4 lay in the tail of their posteriors. We identify likely causes for this with the help of *lalapps_knope* [20].

We demonstrate the method by searching for continuous gravitational wave emission from PSR J1526-2744. The search yields no evidence for a signal, and the obtained 95% confidence upper limits are consistent with those derived with a frequentist method as seen in Fig. 12.

In this paper we assume a simple Gaussian model for the noise, weighting the $F_a$ and $F_b$ quantities according to the estimated noise on a per-SFT basis [31], but we do not account for uncertainties in the noise-level estimation. The other Bayesian known-pulsar search pipeline [20] addresses this issue with an analytical marginalization of the unknown standard deviation of the noise leading to a Student's t-likelihood function. Another approach would be including the uncertainties due to the PSD estimation as additional explicit parameters, and sampling over these with certain priors. A similar approach could also be applied to account for the calibration uncertainties of the detectors, as already done with other types of gravitational wave searches [57,58].

In addition to the targeted application on known pulsars discussed in this paper, a future application of this method is anticipated in the final stages of the follow-up of interesting detection candidates from wide-parameter-space search pipelines. We plan to characterize the method for this application, allowing for additional exploration of (expected small) uncertainties in phase-evolution parameters.

One limitation of this method is an underestimation of $h_0$ in the ultrahigh SNR regime, which is discussed in Appendix B. Full characterization of and potential solutions to this problem lie beyond the scope of this paper and will be considered in future work.

## ACKNOWLEDGMENTS

This research has made use of data or software obtained from the Gravitational Wave Open Science Center [59], a service of LIGO Laboratory, the LIGO Scientific Collaboration, the Virgo Collaboration, and KAGRA. We thank Oliver Behnke for setting up the *knope* pipeline. This project has received funding from the European Union's Horizon 2020 research and innovation program under the Marie Skłodowska-Curie Grant Agreement No. 101029058. This work has utilized the ATLAS computing cluster at the MPI for Gravitational Physics Hannover. We thank the anonymous referee for suggestions that improved the paper.

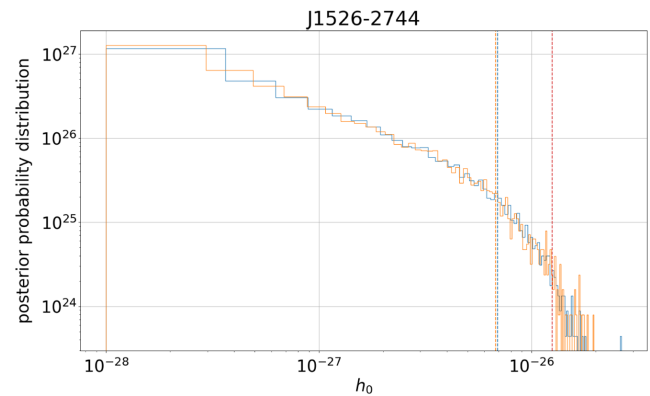

FIG. 12. 95% confidence upper limits on continuous gravitational wave emission from PSR J1526-2744 derived via Bayesian and frequentist methods. Blue: posterior distribution on $h_0$ derived using the full-likelihood method (from Fig. 10), with the 95% upper limit denoted by the dashed-blue line. Orange: posterior distribution on $h_0$ derived using the $\phi_0$ marginalized likelihood method (from Fig. 11), with the 95% upper limit denoted by the dashed-orange line. Dashed-red line: the frequentist upper limit reported in [18].

## APPENDIX A: STATISTICAL PROPERTIES OF THE LOG-LIKELIHOOD RATIO

The log-likelihood ratio $\log\mathcal{L}$ of Eq. (12) depends linearly on the four $x_\mu$, which each follow a Gaussian distribution when the noise is Gaussian. Thus $\log\mathcal{L}$ is also a Gaussian-distributed quantity. In the case of a signal with amplitude parameters $\mathcal{A}_s^\mu$, the four $x_\mu$ have expectation values $s_\mu \equiv E[x_\mu] = \mathcal{A}_s^\nu \mathcal{M}_{\nu\mu}$, and second moment $E[x_\mu x_\nu] = \mathcal{M}_{\mu\nu} + s_\mu s_\nu$. Therefore the expectation of the log-likelihood ratio is

$$\mathrm{E}[\log\mathcal{L}] = \mathcal{A}^\mu s_\mu - \frac{1}{2}\rho^2, \qquad (\mathrm{A}1)$$

which in the perfect-match signal case $\mathcal{A}^\mu = \mathcal{A}_s^\mu$ and the noise case $\mathcal{A}_s^\mu = 0$ yields, respectively,

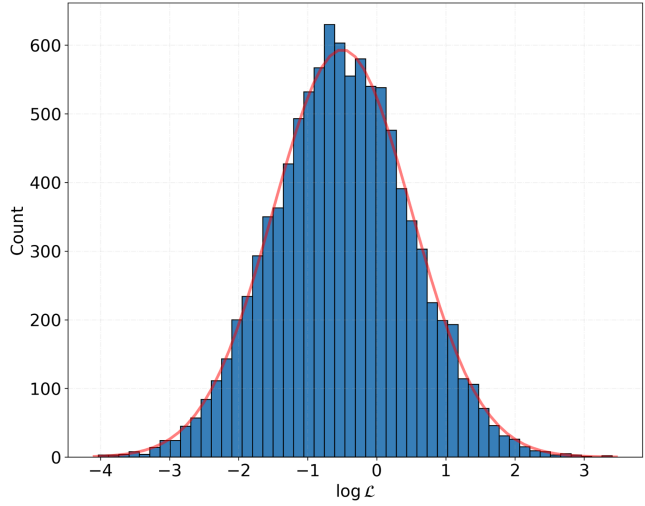

FIG. 13. Histogram of 10 000 log-likelihood ratio $\log \mathcal{L}$ [given by Eq. (12)] values for the noise-only case. The red line shows the expected Gaussian distribution with a mean of $-0.5$ and a standard deviation of 1.

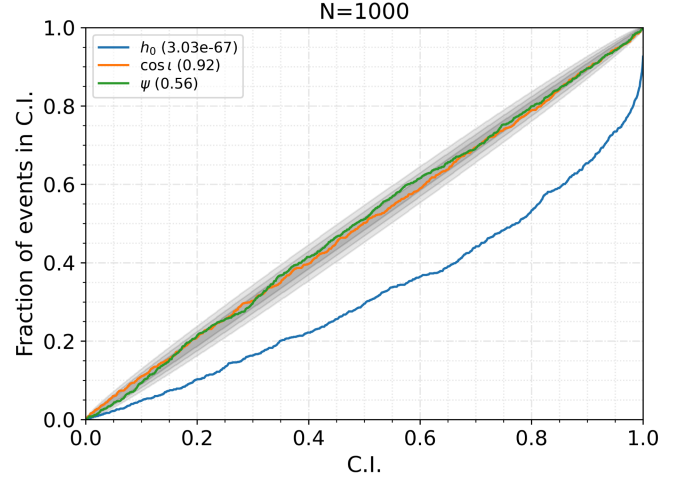

FIG. 14. PP plot of $\phi_0$-marginalized likelihood with a maximum SNR of $\rho \approx 1200$ for the included signals, using the default, $T_{\mathrm{SFT}} = 1800$ s and $\Delta k = 8$ bins. The expected 1-, 2- and 3$\sigma$ deviations under the finite ($N = 1000$) number of injections are in the shaded grey region in decreasing opacity.

$$E[\log \mathcal{L}]_{\mathcal{A}_s = \mathcal{A}} = \frac{1}{2}\rho^2,$$
$$E[\log \mathcal{L}]_{\mathcal{A}_s = 0} = -\frac{1}{2}\rho^2. \quad (A2)$$

The corresponding variance is found as

$$\mathrm{Var}[\log \mathcal{L}] = \mathcal{A}^\mu E[x_\mu x_\nu] \mathcal{A}^\nu - \rho^2 \mathcal{A}^\mu E[x_\mu] + \frac{1}{4}\rho^4 - E[\log \mathcal{L}]^2$$
$$= \rho^2 + (\mathcal{A}^\mu s_\mu)^2 - \rho^2 \mathcal{A}^\mu s_\mu + \frac{1}{4}\rho^4 - E[\log \mathcal{L}]^2$$
$$= \rho^2, \quad (A3)$$

in both the noise and signal cases. Figure 13 shows a histogram with the distribution for the noise-only case, where agreement with the expected Gaussian distribution can be seen.

## APPENDIX B: LIMITATIONS IN THE ULTRAHIGH SNR REGIME

For ultrahigh SNR (of the order of $\rho \approx 1000$) signals, the accuracy of the $h_0$ estimation is compromised.

We set up a PP test with simulated signals of amplitudes drawn from the prior range $h_0 \in [10^{-25}, 10^{-23}]$ in data spanning 10 days with a noise floor of $9 \times 10^{-24}/\sqrt{\mathrm{Hz}}$. The corresponding SNR range of the signals is $\rho \in [12, 1200]$. We use the $\phi_0$-marginalized likelihood and DYNESTY sampler to recover these signals and produce PP plots as described in Sec. III D 2. In the resulting PP plot, shown in Fig. 14, the $h_0$ curve reveals a systematic bias.

The biases in the $h_0$ curve likely arise due to a combination of approximations in the computation of per-SFT quantities contributing to the $\mathcal{F}_a$ and $\mathcal{F}_b$:

(i) the phase evolution of the signal during the time span of an SFT is approximated by a linear term [$\phi(t) = 2\pi f t$] and higher-order corrections are neglected,

(ii) the antenna-pattern coefficients are assumed constant during the time span of an SFT,

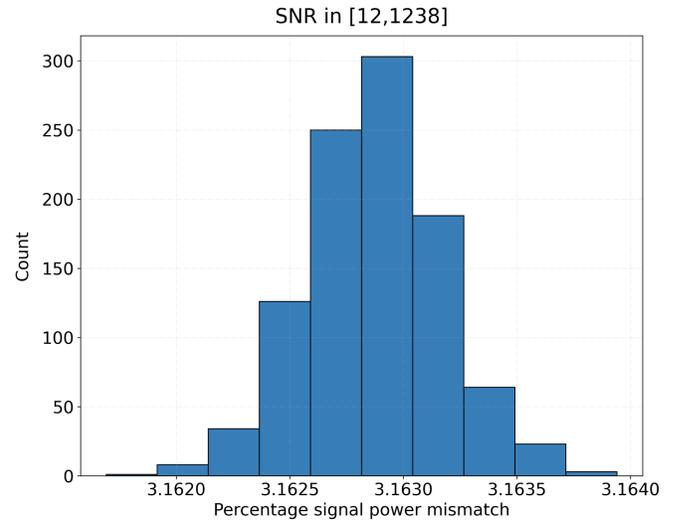

FIG. 15. Percentage mismatch in the signal power ($\rho^2$) of 1000 simulated signals. The approximations in the computations of per-SFT quantities induce an $\approx 3\%$ bias in the posteriors of $h_0$ at $T_{\mathrm{SFT}} = 1800$ s.

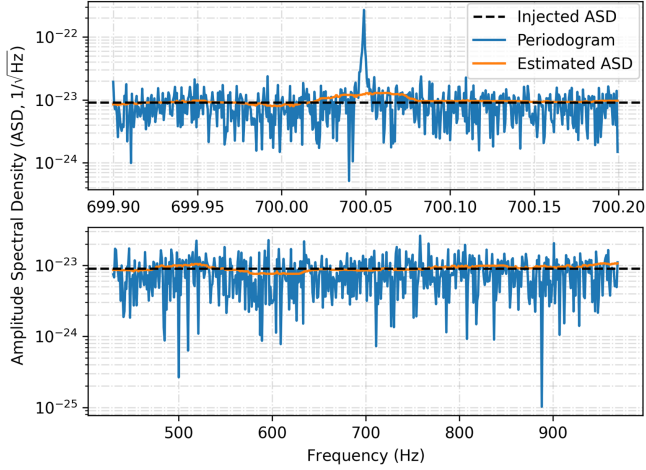

FIG. 16. An example SFT of (i) $T_{SFT} = 1800$ s in the top panel and (ii) $T_{SFT} = 1$ s in the bottom panel, when an ultra-high SNR signal is present. The black-dashed line shows the true value of ASD. In the top panel, the signal appears in the rescaled periodogram ($\sqrt{r}$, blue curve), and the estimated ASD in the SFT (in orange) shows an elevation when the running median window includes the bins elevated by the signal. In the bottom panel, with $T_{SFT} = 1$ s, the per-SFT $\rho^2$ of the signal is lower, and does not affect the noise estimation.

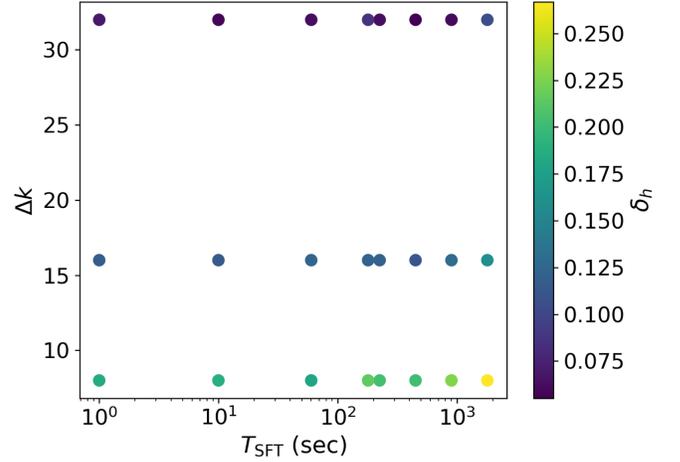

FIG. 17. Effects on the PP plot $h_0$ curve due to the varying time base of SFTs and the number of frequency bins included in the Dirichlet kernel function. In a PP plot, at C.I. = 90% we expect 90% of the simulated signals to be recovered. The color code shows the deviation from this expectation in terms of the difference between the measured value for C.I. = 90% and 90%. The PP plots contain 1000 signals in the SNR range $\rho \in [12, 1200]$ with a frequency of $\approx 700$ Hz. Noise estimation is turned off.

(iii) the number of bins used in the Dirichlet kernel is truncated to a finite number $\Delta k$,
(iv) uncertainties in the noise estimation and biases inherent to the process [60].

To quantify these biases, we simulate 1000 noiseless signals with SNRs in the range $\rho \in [12, 1240]$ and compute the percentage difference in their true signal power and that computed by our codes. For $T_{SFT} = 1800$ s, the bias amounts to $\approx 3\%$ as seen in Fig. 15. In the low-SNR regime, this 3% bias is absorbed in the width of the individual posterior distributions, and its effect does not show up in the PP plots. But in ultrahigh SNR signals, the posterior distributions on the amplitude parameters are narrowly peaked, and systematic biases, even at a few per cent levels, begin to matter. This is seen in the PP plots composed of simulated signals with very high SNR.

It is worth noting that ultrahigh SNR signals can emerge in *individual* SFTs and contaminate the noise floor estimation if their per-SFT signal power ($\rho^2$) is large. For an example SFT (with data $\tilde{x}$) we show the square root of the rescaled periodogram $r$ computed as

$$r = \frac{2}{T_{SFT}} |\tilde{x}|^2, \quad (B1)$$

and the estimated ASD (using a running median with 101 bins) in the top panel of Fig. 16 for $T_{SFT} = 1800$ s and in the bottom panel for $T_{SFT} = 1$ s.

We turn off the noise estimation and explore further the effects of (i), (ii), and (iii) by producing PP plots using the three-dimensional likelihood with varying $T_{SFT}$ and $\Delta k$. We simulate 1000 signals with SNR in the range $\rho \in [12, 1200]$ with phase-evolution parameters of Table II. For every PP plot we compute

$$\delta_h = 0.9 - \text{Coverage}(\text{C.I.} = 0.9) \quad (B2)$$

where Coverage denotes the measured fraction of injections in the credible interval, thus quantifying the bias in the $h_0$ curve. The resulting figure, Fig. 17, shows variation of $\delta_h$ (in color code) with $T_{SFT}$ and $\Delta k$. The $h_0$ bias shows an overall decrease with increasing $\Delta k$ and decreasing $T_{SFT}$.

A study of the interplay between (i), (ii), (iii), and (iv) under different conditions of $T_{SFT}$, $\Delta k$, signal power and phase-evolution parameters is interesting for improving the accuracy of $h_0$ estimation in these SNR regimes, but further study is postponed to future work.

[1] B. Steltner, M. A. Papa, H. B. Eggenstein, R. Prix, M. Bensch, B. Allen, and B. Machenschalk, Astrophys. J. **952**, 55 (2023).
[2] R. Abbott *et al.* (LIGO Scientific, Virgo, and KAGRA Collaborations), Phys. Rev. D **106**, 102008 (2022).
[3] V. Dergachev and M. A. Papa, Phys. Rev. D **109**, 022007 (2024).
[4] V. Dergachev and M. A. Papa, Phys. Rev. X **13**, 021020 (2023).
[5] V. Dergachev and M. A. Papa, Phys. Rev. D **104**, 043003 (2021).
[6] P. B. Covas, M. A. Papa, R. Prix, and B. J. Owen, Astrophys. J. Lett. **929**, L19 (2022).
[7] J. Ming, M. A. Papa, H.-B. Eggenstein, B. Machenschalk, B. Steltner, R. Prix, B. Allen, and O. Behnke, Astrophys. J. **925**, 8 (2022).
[8] R. Abbott *et al.* (LIGO Scientific, Virgo, and KAGRA Collaborations), Phys. Rev. D **106**, 042003 (2022).
[9] R. Abbott *et al.* (LIGO Scientific, Virgo, and KAGRA Collaborations), Phys. Rev. D **106**, 062002 (2022).
[10] M. A. Papa, J. Ming, E. V. Gotthelf, B. Allen, R. Prix, V. Dergachev, H.-B. Eggenstein, A. Singh, and S. J. Zhu, Astrophys. J. **897**, 22 (2020).
[11] R. Abbott *et al.* (LIGO Scientific, Virgo, and KAGRA Collaborations), Astrophys. J. Lett. **941**, L30 (2022).
[12] J. T. Whelan *et al.*, Astrophys. J. **949**, 117 (2023).
[13] Y. Zhang, M. A. Papa, B. Krishnan, and A. L. Watts, Astrophys. J. Lett. **906**, L14 (2021).
[14] R. Abbott *et al.* (LIGO Scientific, Virgo, and KAGRA Collaborations), Astrophys. J. **935**, 1 (2022).
[15] R. Abbott *et al.* (LIGO Scientific, Virgo, and KAGRA Collaborations), Astrophys. J. **932**, 133 (2022).
[16] A. Ashok, B. Beheshtipour, M. A. Papa, P. C. C. Freire, B. Steltner, B. Machenschalk, O. Behnke, B. Allen, and R. Prix, Astrophys. J. **923**, 85 (2021).
[17] L. Nieder *et al.*, Astrophys. J. Lett. **902**, L46 (2020).
[18] C. J. Clark *et al.*, Mon. Not. R. Astron. Soc. **519**, 5590 (2023).
[19] B. P. Abbott *et al.* (LIGO Scientific and Virgo Collaborations), Astrophys. J. **879**, 10 (2019); **899**, 170(E) (2020).
[20] M. Pitkin, M. Isi, J. Veitch, and G. Woan, arXiv:1705.08978.
[21] B. P. Abbott *et al.* (LIGO Scientific and Virgo Collaborations), Astrophys. J. **839**, 12 (2017); **851**, 71(E) (2017).
[22] J. Aasi *et al.* (LIGO Scientific Collaboration), Astrophys. J. **785**, 119 (2014).
[23] B. P. Abbott *et al.* (LIGO Scientific and Virgo Collaborations), Astrophys. J. **713**, 671 (2010).
[24] B. Abbott *et al.* (LIGO Scientific Collaboration), Phys. Rev. D **76**, 042001 (2007).
[25] B. Abbott *et al.* (LIGO Scientific Collaboration), Phys. Rev. Lett. **94**, 181103 (2005).
[26] B. Abbott *et al.* (LIGO Scientific Collaboration), Phys. Rev. D **69**, 082004 (2004).
[27] G. Ashton and R. Prix, Phys. Rev. D **97**, 103020 (2018).
[28] P. Jaranowski, A. Krolak, and B. F. Schutz, Phys. Rev. D **58**, 063001 (1998).
[29] R. Prix and B. Krishnan, Classical Quantum Gravity **26**, 204013 (2009).
[30] C. Cutler and B. F. Schutz, Phys. Rev. D **72**, 063006 (2005).
[31] R. Prix, The F-statistic and its implementation in ComputeFstatistic_v2, Technical Report LIGO-T0900149, LIGO, 2018, https://dcc.ligo.org/LIGO-T0900149/public.
[32] L. S. Finn, Phys. Rev. D **46**, 5236 (1992).
[33] LIGO Scientific Collaboration, LALSuite: LIGO Scientific Collaboration Algorithm Library Suite, Astrophysics Source Code Library, record ascl:2012.021 (2020).
[34] R. J. Dupuis and G. Woan, Phys. Rev. D **72**, 102002 (2005).
[35] K. Riles, Living Rev. Relativity **26**, 3 (2023).
[36] G. Pagliaro, M. A. Papa, J. Ming, J. Lian, D. Tsuna, C. Maraston, and D. Thomas, Astrophys. J. **952**, 123 (2023).
[37] M. Isi, M. Pitkin, and A. J. Weinstein, Phys. Rev. D **96**, 042001 (2017).
[38] P. R. Williams and B. F. Schutz, AIP Conf. Proc. **523**, 473 (2000).
[39] A. Gelman, J. B. Carlin, H. S. Stern, D. B. Dunson, A. Vehtari, and D. B. Rubin, *Bayesian Data Analysis* (Chapman and Hall/CRC, 2020).
[40] G. Ashton *et al.*, Astrophys. J. Suppl. Ser. **241**, 27 (2019).
[41] K. Wette, SoftwareX **12**, 100634 (2020).
[42] J. S. Speagle, Mon. Not. R. Astron. Soc. **493**, 3132 (2020).
[43] G. Ashton *et al.*, Nature (London) **2**, 39 (2022).
[44] D. Yallup, T. Janßen, S. Schumann, and W. Handley, Eur. Phys. J. C **82**, 8 (2022).
[45] G. Ashton and C. Talbot, Mon. Not. R. Astron. Soc. **507**, 2037 (2021).
[46] J. T. Whelan, R. Prix, C. J. Cutler, and J. L. Willis, Classical Quantum Gravity **31**, 065002 (2014).
[47] M. Abramowitz and I. A. Stegun, *Handbook of Mathematical Functions*, 9th ed. (Dover Publications, New York, 1964).
[48] R. Abbott *et al.* (LIGO Scientific, Virgo, and KAGRA Collaborations), Phys. Rev. D **104**, 082004 (2021).
[49] M. Pitkin, C. Gill, D. I. Jones, G. Woan, and G. S. Davies, Mon. Not. R. Astron. Soc. **453**, 4399 (2015).
[50] R. Abbott *et al.* (LIGO Scientific and Virgo Collaborations), SoftwareX **13**, 100658 (2021).
[51] B. Steltner, M. A. Papa, and H.-B. Eggenstein, Phys. Rev. D **105**, 022005 (2022).
[52] G. Woan, M. D. Pitkin, B. Haskell, D. I. Jones, and P. D. Lasky, Astrophys. J. Lett. **863**, L40 (2018).
[53] B. J. Owen, Phys. Rev. Lett. **95**, 211101 (2005).
[54] C. Rover, C. Messenger, and R. Prix, in *PHYSTAT 2011* (CERN, Geneva, 2011), pp. 158–163.
[55] R. Abbott *et al.* (LIGO Scientific and Virgo Collaborations), Astrophys. J. Lett. **902**, L21 (2020).
[56] M. Pitkin, J. Open Source Softwaare **7**, 4568 (2022).
[57] S. Vitale, C.-J. Haster, L. Sun, B. Farr, E. Goetz, J. Kissel, and C. Cahillane, Phys. Rev. D **103**, 063016 (2021).
[58] E. Payne, C. Talbot, P. D. Lasky, E. Thrane, and J. S. Kissel, Phys. Rev. D **102**, 122004 (2020).
[59] www.gw-openscience.org.
[60] R. Prix, F-statistic bias due to noise-estimator, Technical Report LIGO-T1100551-v1, LIGO, 2006, https://dcc.ligo.org/LIGO-T1100551/public.